\magnification=1200

\headline{\ifnum\pageno=1 \nopagenumbers
\else \hss\number \pageno \fi}

\overfullrule=0pt
\footline={\hfil}
\font\boldgreek=cmmib10
\textfont9=\boldgreek
\mathchardef\mypsi="0920

\mathchardef\myphi="091E

\def\lsim{\raise0.3ex\hbox{$<$\kern-0.75em\raise-1.1ex\hbox{$\sim$}}}
\def\gsim{\raise0.3ex\hbox{$>$\kern-0.75em\raise-1.1ex\hbox{$\sim$}}}
\baselineskip=10pt
\vbox to 1,5truecm{}
\parskip=0.2truecm
\centerline{\bf STRUCTURE FUNCTIONS OF NUCLEI AT SMALL x} \smallskip
\centerline{\bf AND DIFFRACTION AT HERA}\bigskip

\bigskip \centerline{by}\medskip
\centerline{{\bf A. Capella, A. Kaidalov}\footnote{*}{Permanent address : ITEP,
B. Cheremushkinskaya 25, 117259 Moscow, Russia}{\bf, C. Merino}\footnote{**}
{Permanent address : Universidade Santiago de Compostela, Dep. F{\'\i}sica de
Particulas, E-15706 Santiago de Compostela, Spain}{\bf, D. Pertermann}
\footnote{***}{Permanent address : Univ-GH-Siegen, Phys. Dept., D-57068 Siegen,
Germany} {\bf and J. Tran Thanh Van}}
\smallskip
 \centerline{Laboratoire de Physique Th\'eorique et Hautes Energies
\footnote{****}{Laboratoire associ\'e au Centre National de la Recherche
Scientifique - URA D0063}}  \centerline{Universit\'e de Paris XI,
b\^atiment 211, 91405 Orsay cedex, France}
\bigskip \bigskip \bigskip\baselineskip=20pt
\noindent
${\bf Abstract}$ \par
Gribov theory is applied to investigate the shadowing effects in the structure
functions of nuclei. In this approach these effects are related to the process
of diffractive dissociation of a virtual photon. A model for this diffractive
process, which describes well the HERA data, is used to calculate the shadowing
in nuclear structure functions. A reasonable description of the $x$, $Q^2$ and
$A$-dependence of nuclear shadowing is achieved. \par

\vbox to 4 truecm{}

\noindent LPTHE Orsay 97-07 \par
\noindent July 1997
\vfill \supereject
\noindent {\bf 1. \underbar{Introduction}} \par \vskip 5 truemm
Deep inelastic scattering (DIS) on nuclei gives important information on
distributions of quarks and gluons in nuclei. The region of small Bjorken $x$
is especially interesting because partonic clouds of different nucleons overlap
as $x \to 0$ and shadowing effects become important. There are experimental
results in this region, which show that there are strong deviations from an
$A^1$ behavior in the structure functions [1]. Several theoretical models have
been proposed to understand these data [1]. The most general approach is based
on the Gribov theory [2]. It relates partonic and hadronic descriptions of
small $x$ phenomena in interactions of real or virtual photons with nuclei. In
this approach the shadowing effects can be expressed in terms of the
cross-sections for diffraction dissociation of a photon on a nucleon (Fig. 1).
This process has been studied recently in DIS at HERA [3]. The detailed $x$,
$Q^2$ and $M^2$ ($M$ is the invariant mass of the diffractively produced system)
dependencies observed in these experiments have been well described in the
theoretical model of ref. [4] which is based on Regge factorizations and uses
as an input available information on diffractive production in hadronic
interactions. Here we will apply the same model to calculate the structure
functions of nuclei in the small $x$-region. The use of the model, which
describes well the diffraction dissociation of virtual photons on a nucleon
target, leads to a strong reduction of the theoretical uncertainty in
calculations of the structure functions of nuclei in comparison with previous
calculations [1, 5-8]. It also allows to discuss the shadowing effects in gluon
distributions.
 \par \vskip 5 truemm

\noindent {\bf 2. \underbar{The model}} \par \vskip 5 truemm

In the Gribov approach the forward scattering amplitude of a photon with
virtuality $Q^2$ on a nuclear target can be written as the sum of the
diagrams shown in Fig. 2. Since we are interested in the low $x$ region we will describe
the various $\gamma^*N$ interactions by Pomeron exchange. The diagram of Fig. 2a corresponds
to the sum of interactions with individual nucleons and is propotional to $A^1$. The second
diagram (2b) contains a double scattering with two target nucleons. It gives a negative
contribution to the total cross-section, proportional to $A^{4/3}$ (for large $A$).
It describes the first shadowing correction for sea quarks. According to reggeon diagram
technique [9] and Abramovsky, Gribov, Kancheli (AGK) cutting rules [10], the contribution of
the diagram of Fig. 2b to the total $\gamma^*A$ cross-section is related to the diffractive
production of hadrons by a virtual photon as follows~: $$\sigma_A^{(2)} = - 4 \pi \ A(A - 1)
\int d^2b \ T_A^2(b) \int_{M^2_{min}}^{M^2_{max}} dM^2 \left . {d \sigma_{\gamma^*p}^{\cal D}
\over dM^2dt} \right |_{t=0} F_A^2(t_{min}) \ \ \ , \eqno(1)$$ 
\noindent where $T_A(b)$ is the nuclear profile
function, $\rho_A$ is the nuclear density
($T_A(b) = \int_{- \infty}^{+\infty} dZ
\rho_A(b, Z),\break \noindent \int d^2b \ T_A(b) = 1$) and
$$F_A(t_{min}) = \int d^2b J_0 (\sqrt{- t_{min}} b) T_A(b) \ ,
\ t_{min} =  - m_N^2 \ x^2
\left ( {Q^2 \over M^2 + Q^2} \right )^{- 2} \ \ \
.$$

\noindent Note that
$F_A(t_{min})$ is equal to unity as $x \to 0$ and decreases fast as $x$
increases to $x_{cr} \sim {1 \over m_N R_A}$, due to a lack of coherence for
$x > x_{cr}$. \par

Eq. (1) is written in the approximation $R_A^2 \gg R_N^2$, where $R_N$ is the
radius of the $\gamma^*p$ interaction. It will be used in this form only for $A >
20$ (see below). We have also neglected the real part of the Pomeron
amplitude which is small for our value of the Pomeron intercept (see Eq. (5)). However,
for higher values of this intercept the contribution of the real part can be substantial
[11]. \par

For a deuteron, the double rescattering contribution has the following form

$$\sigma_D^{(2)} = - 2 \int_{- \infty}^{t_{min}} dt \int_{M^2_{min}}^{M^2_{max}} dM^2 {d
\sigma^{\cal D}_{\gamma^*N} \over dM^2 dt} F_D(t) \eqno(2)$$

\noindent where $F_D(t) = \exp (at)$, with $a = 40$ GeV$^{-2}$. $M_{min}^2$
in eqs. (1), (2) corresponds to the minimal mass of the diffractively produced
hadronic system and $M_{max}^2$ is chosen according to the condition~: $x_P = x
\cdot {M^2 + Q^2 \over Q^2} \simeq {M^2 + Q^2 \over W^2} \leq 0.1$. \par

Equation (2) has been used to
calculate inelastic contributions to Glauber corrections in hadron-deuteron interactions [12,
13] and was generalized to heavier nuclei in the form (1) in ref. [14]. \par

Thus the second order rescattering term can be calculated if the differential
cross-section for diffractive production by a virtual photon is known. \par

Higher order rescatterings are model dependent, but calculation shows that, for the values
of $A$ and $x$ $(x \ \gsim \ 10^{-3})$ where experimental data exist, their contribution is
rather small. We use the following unitary expression for the total
$\gamma^*A$ cross-section $$\sigma_{\gamma^*A} = \sigma_{\gamma^*N} \int d^2b {A \ T_A(b)
\over 1 + (A - 1) f(x,Q^2) T_A(b)} \eqno(3)$$ \noindent where $$f(x, Q^2) = 4 \pi \int dM^2
\left .
{d\sigma_{\gamma^*p}^{\cal D} \over dM^2dt} \right
|_{t=0} F_A^2(t_{min})/\sigma_{\gamma^*N} \ \ \ . $$
\noindent This expression is valid in the generalized Schwimmer model [15, 16]
and is obtained from a summation of fan diagrams
with triple Pomeron interaction. However, its physical basis and applicability is much
broader. For example it follows from the rescattering of a $q\bar{q}$ system with transverse
sizes distributed according to a gaussian [17]. We have checked that the results obtained
from a summation of higher order rescatterings of an eikonal type are very similar to the ones
obtained with (3) - the differences being of the order of one percent. \par

Thus we have for the ratio $R_A = F_{2A}/F_{2N}$ of nucleus and nucleon structure functions,
in the region of small $x$ $${F_{2A} \over F_{2N}} = \int d^2b {A \ T_A(b)
\over 1 + (A - 1) f(x, Q^2) T_A(b)} \ \
\ . \eqno(4)$$
\noindent The deviation of this ratio from $A^1 = A \int d^2b T_A(b)$ is due to
the second term in the denominator of the integrand in eq. (4). Thus, knowing
the differential cross-section for diffraction dissociation on a nucleon and
the structure function of a nucleon ($\sigma_{\gamma^*N})$, one can predict the
$A$ (and $x$, $Q^2$) dependence of structure functions of nuclei. Eq. (4) can
only be used in the region of $x$ substantially smaller than $10^{-1}$ where the sea quarks
component dominates. For $x$ close to $10^{-1}$ shadowing of valence quarks (which
in general is not described by eq. (4)) becomes important [18, 19]. The effects
leading to antishadowing (such as real parts in the rescattering diagram due
to secondary exchanges) are also important in the region of $x \sim 0.1$. \par

In refs. [4] we described the diffractive contribution to DIS in terms of
Pomeron exchange
$$F_2^{\cal D}(x, Q^2, x_P, t) = {(g_{pp}^P(t) )^2 \over 16 \pi}
x_P^{1-2\alpha_P(t)} F_P(\beta, Q^2, t) \eqno(5)$$
\noindent where $g_{pp}^P(t)$ is the Pomeron-proton coupling $(g_{pp}^P(t) =
g_{pp}^P(0) \exp (Ct)$ with $(g_{pp}^P(0))^2 =$ 23~mb and
$C = 2.2$ GeV$^{-2}$), $\alpha_P(t) = \alpha_P(0) + \alpha '_P(0)t$ is the Pomeron trajectory
($\alpha_P(0)$ = 1.13, $\alpha '_P(0) = 0.25$ GeV$^{-2}$) and
$F_P(\beta , Q^2, t)$ is the Pomeron structure function. The variable
$\beta = {Q^2 \over M^2 + Q^2} = {x \over x_P}$ plays the same role for the
Pomeron as the Bjorken variable $x$ for the proton. At large $Q^2$, $F_P$ can
be expressed in terms of the quark distributions in the Pomeron
$$F_P(\beta , Q^2,t) = \sum_i e_i^2 \beta \left [ q_i^P(\beta , Q^2, t) +
\bar{q}_i^P(\beta , Q^2, t) \right ] \ \ \ .  \eqno(6)$$
\noindent In refs. [4] we determined $F_P(\beta , Q^2,t)$ using
Regge-factorization for small values of $\beta$ and a plausible assumption on
the $\beta \to 1$ behavior.
This function was then used as an initial condition for QCD evolution of
partons in the Pomeron. The results of the QCD-evolution crucially depend on
the form of the gluon distribution in the Pomeron. Experimental results for
$F_2^{\cal D}$ can be understood only if the distribution of gluons in the
Pomeron is rather hard and the gluons carry the main part of the Pomeron
momentum [4, 20-22]. The explicit forms of all these functions are given in Appendix 1. \par

The validity of Pomeron factorization (5) for $F_2^{\cal D}$ as well as that of the QCD
evolution for partons in the Pomeron has been questioned in recent papers. These
papers deal with diffractive charm production [23, 24] and with the contribution
of longitudinal photons to diffractive production [25, 26]. However, in all
these papers high-twist effects (in $Q^2$ or $M^2_Q$, where $M_Q$ is the mass of the heavy
quark), which give small contributions to diffractive cross-sections, were considered.
Arguments in favour of usual QCD evolution for the main twist contribution to $F_2^{\cal D}$
have been given in ref. [27]. In any case the CKMT model [4] gives a reasonable
description of diffractive production in DIS. Thus it effectively includes high
twist effects and can be used to compute the function $f(x, Q^2)$, which
determines the shadowing of nuclear structure function via (4). This function
can be written in terms of the ratio $F_P/F_{2N}$~: $$f(x, Q^2) = \int {d\beta
\over 4 \beta} \left ( g_{pp}^P(0) \right )^2 \left ( {1 \over x_P} \right
)^{2\Delta} {F_P(\beta , Q^2) \over F_{2N} (x , Q^2)} F^2_A(t_{min}) 
\eqno(7)$$ 

\noindent where the integration limits are $x/x_{0P}$ with $x_{0P} = 0.1$ and
$Q^2/(M_{min}^2 + Q^2)$. In the following we take $M_{min}^2 =$ 0.4 GeV$^2$, in order to
include the $\rho$-meson peak in the integration region. \par

The parametrization of the Pomeron [4] and nucleon [28] structure functions are given in
Appendix 1. Note that the $Q^2$-dependence of nuclear shadowing is obtained by
evolving separately the nucleon and Pomeron structure functions and taking their
ratio in eq. (7). Actually, one should compute first $F_{2A}$ at $Q^2 = Q_0^2$
and evolve it using the nuclear partonic distributions. However, this would
require the knowledge of these distributions for all values of $x$. At small
$x$, where sea quarks are dominant, these two procedures are equivalent for the
Born term and the first rescattering correction in eq. (4). As discussed above,
higher rescattering corrections are small. \par

In the numerical calculations we use a standard Woods-Saxon profile $T_A(b)$
for $A > 20$. For light nuclei ($A < 20$) we use a gaussian profile

$$T_A(b) = {3 \over 2 \pi R_A^2} \exp (- 3b^2/2 R_A^2) \eqno(8)$$

\noindent with an r.m.s. radius parametrized as [29]

$$R_A = 0.82 \ A^{1/3} + 0.58 \ {\rm fm} \ \ \ . \eqno(9)$$

\noindent For deuteron eq. (9) is not valid. In this case we use eq. (2). The
simple exponential form of $F_D(t)$ gives results which differ by less than
20~$\%$ from the more sophisticated parametrization used in refs. [6] [30]. \par

In eq. (1) we have neglected the $t$-dependence of the $\gamma^*p$
diffractive cross-section. As explained above, this approximation is only used
for large nuclei where nucleon sizes can be neglected as compared to nuclear
ones. For light nuclei ($A < 20$), we take into account this $t$ (or
$b$)-dependence by making the following replacement

$$R_A^2 \Rightarrow R_A^2 + R_N^2 \quad , \qquad R_N = 0.8 \ {\rm fm} \ \ \ .
\eqno(10)$$

\noindent This nucleon radius approximately describes the $t$-dependence of the
$\gamma^*p$ diffractive cross-section in the kinematical region we are
interested in.

\par \vskip 5 truemm

\noindent {\bf 3. \underbar{Numerical results}} \par \vskip 5 truemm

The results of our calculations are shown in Figs. 3-8. Theoretical predictions
for the deuteron structure function $F_{2}^D/2F_2^N$ are shown in Fig. 5. Our results
are close to those of refs. [30, 31] but smaller by a factor of about 3 from
the results of ref. [6]. Comparison of our predictions for the ratio ${2 \over
A} F_2^A/F_2^D$ with experimental data of NMC [32] is shown in Fig. 3 and for
ratios of different nuclei in Fig. 4. New data for the ratio $F_2^{S_n}/F_2^C$
[33] are also shown in Fig. 4. It is important to note that experimental points
in Figs. 3, 4 for different $x$ correspond to different values of $<Q^2>$ [32]
[33]. This correlation has been taken into account in our calculations. The
agreement between theoretical predictions and experimental data is good. Note that 
there are no free parameters in our calculations. \par \vfill \supereject

Our predictions for the ratio ${1 \over A} {F_{2A} \over F_{2N}}$ in the region
of very small $x$ are shown in Figs. 5 for fixed values of $Q^2$. They could be
confronted to experiment if nuclei at HERA would be available. Note that our
results are more reliable for small values of $x$ ($x < 10^{-2}$) where the
effects of both valence quark shadowing and antishadowing are negligeable. The
curves for shadowing effects in the gluon distribution of nuclei ${1 \over A}
{g^A(x, Q^2) \over g^N(x,Q^2)}$ are shown in Figs. 6. They look similar to the
shadowing for the quark case. However the absolute magnitude of the shadowing
is smaller in the gluon case contrary to expectations of some theoretical
models [1] but in agreement with [34]. (Note that these results are sensitive
to the gluonic distribution in the Pomeron, which is poorly known at present).
These predictions can be tested in experimental studies of $J/\psi$ and
$\Upsilon$-production on nuclear targets at RHIC and LHC. \par

Finally we want to discuss in more detail the $Q^2$-dependence of the
shadowing. Recent NMC data [33] for the ratio of $F_2^{S_n}/F_2^C$ are shown in
Figs. 7 as functions of $Q^2$ for fixed values of $x$ in the small $x$ region.
The theoretical curves have a weak dependence on $Q^2$ and are in reasonable
agreement with experiment, although the $Q^2$ dependence seems stronger in the
data especially in the region of small $Q^2$. At larger values of $x$ the data
are practically $Q^2$-independent. These properties should be checked in future
experiments. \par
\vskip 5 truemm

\noindent {\bf 4. \underbar{Conclusions}} \par \vskip 5 truemm

In conclusion, a model based on the Gribov-Glauber theory of nuclear shadowing
and the properties of diffraction in DIS observed at HERA, leads to a fair
description of experimental data on structure functions of nuclei in the small
$x$ region. Predictions of shadowing effects for quark and gluon distributions
are given. They can be tested in future experiments at HERA and in hadronic
colliders. \par \vfill \supereject

\noindent {\bf Acknowledgements} \par
One of us (A.C.) would like to thank G. Do Dang for discussions. A. K. and
D. P. wish to thank the LPTHE for hospitality during a period when this work
was initiated. This work has been partially supported by grant 93-0079 of INTAS.
A. K. also acknowledges support from the grant N$^{\circ}$ 96-02-19184 of RFFI.
\vfill \supereject
\centerline{\bf Appendix 1} \vskip 3 truemm
For the Pomeron and nucleon structure functions we use the parametrization of the CKMT model
[4, 28]
$$F_{2N}(x, Q^2) = A(Q^2) x^{- \Delta (Q^2)} (1 - x)^{n(Q^2) + 4} + B(x, Q^2) x^{1 -
\alpha_R(0)} (1 - x)^{n(Q^2)} \ , \eqno(A.1)$$
$$F_P(\beta , Q^2) = F_{2N} (\beta , Q^2; A \to eA , B(x) \to fB', n\to n-2) \eqno(A.2)$$
\noindent where
$$A(Q^2) = A \left ( {Q^2 \over Q^2 + a} \right )^{1 + \Delta (Q^2)} \ \ , \ \ B(x, Q^2) =
B(x) \left ( {Q^2 \over Q^2 + b} \right )^{\alpha_R}$$
$$\Delta (Q^2) = \Delta_0 \left ( 1 + {2Q^2 \over Q^2 + d} \right ) \ \ , \ \ n(Q^2) = {3
\over 2} \left ( 1 + {Q^2 \over Q^2 + c} \right )$$
\noindent with (all dimensional quantities are in Gev$^2$)
$$A = 0.1502 \ \ , \ \ B' = 1.2035 \ \ , \ \ \alpha_R = 0.4150 \ \ , \ \ \Delta_0 = 0.07684$$
$$a = 0.2631 \ \ , \ \ b = 0.6452 \ \ c = 3.5489 \ \ , \ \ d = 1.1170 \ \ , \ \ e = f = 0.07
\ \ \ .$$
\noindent Finally, we have [28]
$$B(x) = 0.754 + 0.4495 (1 - x) \ \ \ . \eqno(A.3)$$
\noindent The two terms in Eq. (A.3) appear because we have used for the valence quark
distributions in the proton $d(x) = u(x) (1 - x)$. Such a difference between $u$ and $d$
quark distributions is not present in the Pomeron case, and we have dropped the $1 - x$
factor in (A.3). In the CKMT model the Pomeron structure function is determined from the
nucleon one using triple Regge couplings determined from soft diffraction data and Regge
factorization. \par

Comparison of eqs. (A.2) and (6) allows to determine the valence and sea quark
distributions in the Pomeron. Likewise one can determine the corresponding ones in the
nucleon. These distributions are used as initial conditions at $Q^2 = Q_0^2$ in the DGLAP
evolution equation, as described in [28]. The gluon distributions for $Q^2 \leq Q_0^2$ in
the nucleon and the Pomeron are [4, 28]
$$g^N(x, Q^2) = A_g(Q^2) \ x^{-\Delta (Q^2)} (1 - x)^{n(Q^2)+2} \eqno(A.4)$$
$$g_{_P}(x, Q^2) = e \ A_g(Q^2) \ x^{- \Delta (Q^2)} (1 - x)^{-0.5} \eqno(A.5)$$
\noindent where $A_g(Q^2)$ has the same form as $A(Q^2)$ in eqs. (A.1) and (A.2)~:
$$A_g(Q^2) = A_g \left ( {Q^2 \over Q^2 + a} \right )^{1 + \Delta (Q^2)} \ \ \ .$$

The normalization of $g^N$ is obtained from the energy-momentum sum-rule. For the Pomeron
this sum-rule is not valid and the normalization of $g_P$ is obtained from that of $g^N$ using
Regge factorization. The constant $e = 0.07$ is the same as in eq. (A.2). Actually there is a
large ambiguity in the shape of $g_P$. We have only determined the $x$-behaviour at small $x$ as well as
the absolute normalization. The form (A.5) is just a simple extrapolation to the
region of $x \to 1$. \par

Using the gluon distributions in eqs. (4) and (7) one can obtain the
cor\-res\-pon\-ding distributions in nuclei. \par

In order to have exactly the diffractive cross-section computed in ref. [4] as well as the
same $F_{2N}$ of ref. [28] we use the values of $Q^2_0$ in those references. These are
$Q_0^2 =$ 5 GeV$^2$ for the Pomeron and $Q_0^2 =$ 2 GeV$^2$ for the nucleon. The
corresponding gluon normalizations obtained from the energy-momentum sum rule are $A_g =
1.84$ at $Q_0^2$ = 2~GeV$^2$ and $A_g =$ 1.71 at $Q_0^2 =$ 5 GeV$^2$.

\vfill\supereject \centerline{\bf References} \vskip 3 truemm
\item{[1]} M. Arneodo, Phys. Reports {\bf 240} (1994) 301 (and references
therein).
\item{[2]} V. N. Gribov, ZhETF {\bf 56} (1969) 892, ibid {\bf 57} (1969) 1306
[Sov. Phys. JETP {\bf 29} (1969) 483, {\bf 30} (1970) 709].
\item{[3]} T. Ahmed et al (H1 collaboration), Phys. Lett. {\bf B348} (1995) 681.
 \item{} M. Derrick et al (Zeus collaboration), Z. Phys. {\bf C68} (1995) 569~;
         Z. Phys. {\bf C70} (1996) 391.
 \item{[4]} A. Capella, A. Kaidalov, C. Merino and J. Tran Thanh Van, Phys.
               Lett. {\bf B343} (1995) 403.
\item{} A. Capella, A. Kaidalov, C. Merino, D. Pertermann and J. Tran Thanh
Van, Phys. Rev. {\bf D53} (1996) 2309. \item{[5]} K. Boreskov, A.
Capella, A. Kaidalov and J. Tran Thanh Van, Phys. Rev. {\bf D47} (1993) 219.
\item{[6]} V. Barone et al, Z. Phys. {\bf C58} (1993) 541.
\item{[7]} W. Melnitchouk, A. W. Thomas, Phys. lett. {\bf B317} (1993) 437~; Phys. Rev.
{\bf C52} (1995) 3373.
\item{[8]} G. Piller, W. Ratzka, W. Weise, Z. Phys. {\bf A352} (1995) 427. 
\item{[9]} V. N. Gribov, ZhETF {\bf 57} (1967) 654 [Sov. Phys. JETP {\bf 26}
           (1968) 14].
\item{[10]} V. A. Abramovsky, V. N. Gribov and O. V. Kancheli, Yad. Fiz.
           {\bf 18} (1973) 595 [Sov. J. Nucl. Phys. {\bf 18} (1974) 308].
\item{[11]} A. Bialas, W. Czyz and W. Florkowski, TPJU-25/96 and references therein. 
\item{[12]} A. B. Kaidalov, L. A. Kondratyuk, JETP Letters {\bf 15} (1972) 170~; Nucl. Phys.
{\bf B56} (1973) 90.

\item{[13]} V. V. Anisovich, L. G. Dakhno and P. E. Volkovitsky, Yad. Fiz. {\bf 15} (1972)
168 [Sov. J. Nucl. Phys. {\bf 15} (1972) 97].
\item{[14]} V. A. Karmanov and L. A. Kondratyuk, JETP Letters {\bf 18} (1973) 266.
 \item{[15]} A. Schwimmer, Nucl. Phys. {\bf B94}
(1975) 445. 
\item{[16]} K. G. Boreskov et al., Yad. Fiz. {\bf 53} (1991) 569 [Sov. J. Nucl.
           Phys. {\bf 53} (1991) 356].
\item{[17]} B. Kopeliovich et al., JINR E2-86-125.
\item{[18]} L. L. Frankfurt, M. I. Strikman and S. Liuti, Phys. Rev. Lett.
            {\bf 65} (1990) 1725.
\item{[19]} A. B. Kaidalov, C. Rasinariu and U. Sukhatme, UICHEP-TH/96-9.
\item{[20]} T. Gehrmann and W. J. Stirling, Z. Phys. {\bf C70} (1996) 89.
\item{[21]} K. Golec-Biernat and J. Kwiecinski, Phys. Lett. {\bf B353} (1995)
            329.
\item{[22]} J.
Dainton (H1 collaboration), Proceedings Workshop on Deep Inelastic Scattering
and QCD, Paris, France, 24-28 April 1995 (ed. J. P. Laporte and Y Sirois).
\item{[23]} M. Genovese, N. N. Nikolaev, B. G. Zakharov, Phys. Lett. {\bf B378} (1996) 347.
\item{[24]} E. M. Levin, A. D. Martin, M. G. Ryskin, T. Teubner, DTP 96-50.
\item{[25]} J. Bartels, H. Lotter, M. Wusthoff, Phys. Lett. {\bf B379} (1996) 248~; J.
Bartels, C. Ewerz, H. Lotter, M. Wusthoff, Phys. Lett. {\bf B386} (1996) 389.
\item{[26]} M. Genovese, N. N. Nikolaev, B. G. Zakharov, Phys. Lett. {\bf B380} (1996) 213.
\item{[27]} A. Berera and D. E. Soper, Phys. Rev. {\bf D53} (1996) 6162.
\item{[28]} A. Capella, A. Kaidalov, C. Merino, J. Tran Thanh Van, Phys. Lett
            {\bf B337} (1994) 358.
\item{[29]} M. A. Preston and R. K. Bhoduri, Structure of the Nucleus,
            Addison-Wesley, New York 1975. 
\item{[30]} W. Melnitchoak, A. W. Thomas, Phys. Rev. {\bf D47} (1993) 3783.
\item{[31]} B. Badelek, J. Kwiecinski, Phys. Rev. {\bf D50} (1994) 4.
\item{[32]} P. Amandruz et al
            (NMC collaboration), Nucl. Phys. {\bf B441} (1995) 3.
\item{[33]} N. Arneodo et al (NMC collaboration), Nucl. Phys. {\bf
            B481} (1996) 23.
\item{[34]} K. J. Eskola, Nucl. Phys. {\bf B400} (1994) 240.

\vfill \supereject
\centerline{\bf Figure Captions}
\vskip 3 truemm
{\parindent= 2 truecm
\item{\bf Fig. 1 :} Diffractive dissociation of a virtual photon. The shaded
     area represents the exchange of a Pomeron.
\vskip 3 truemm

\item{\bf Fig. 2 :} The first two terms (single and double scattering) of the
multiple scattering series for the total $\gamma^*N$ cross-section in the
Gribov-Glauber approach.
\vskip 3 truemm

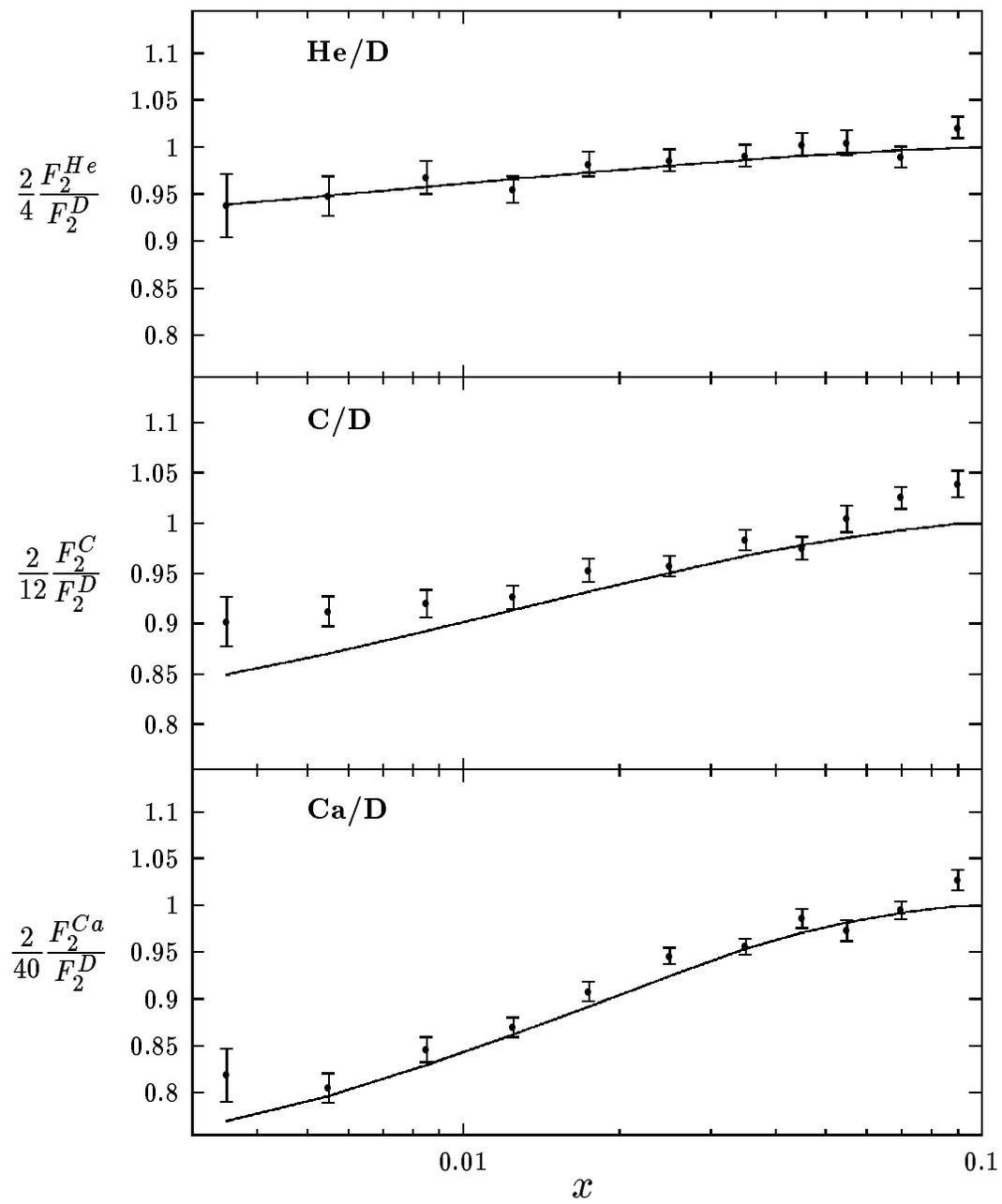
\item{\bf Fig. 3 :} The ratios $(2/A) F_2^A/F_2^D$ computed from eq. (3) for
     different values of $x$. The experimental points are from ref. [17]. The
     values of $Q^2$ are different for different$x$-values [17].

\vskip 3 truemm
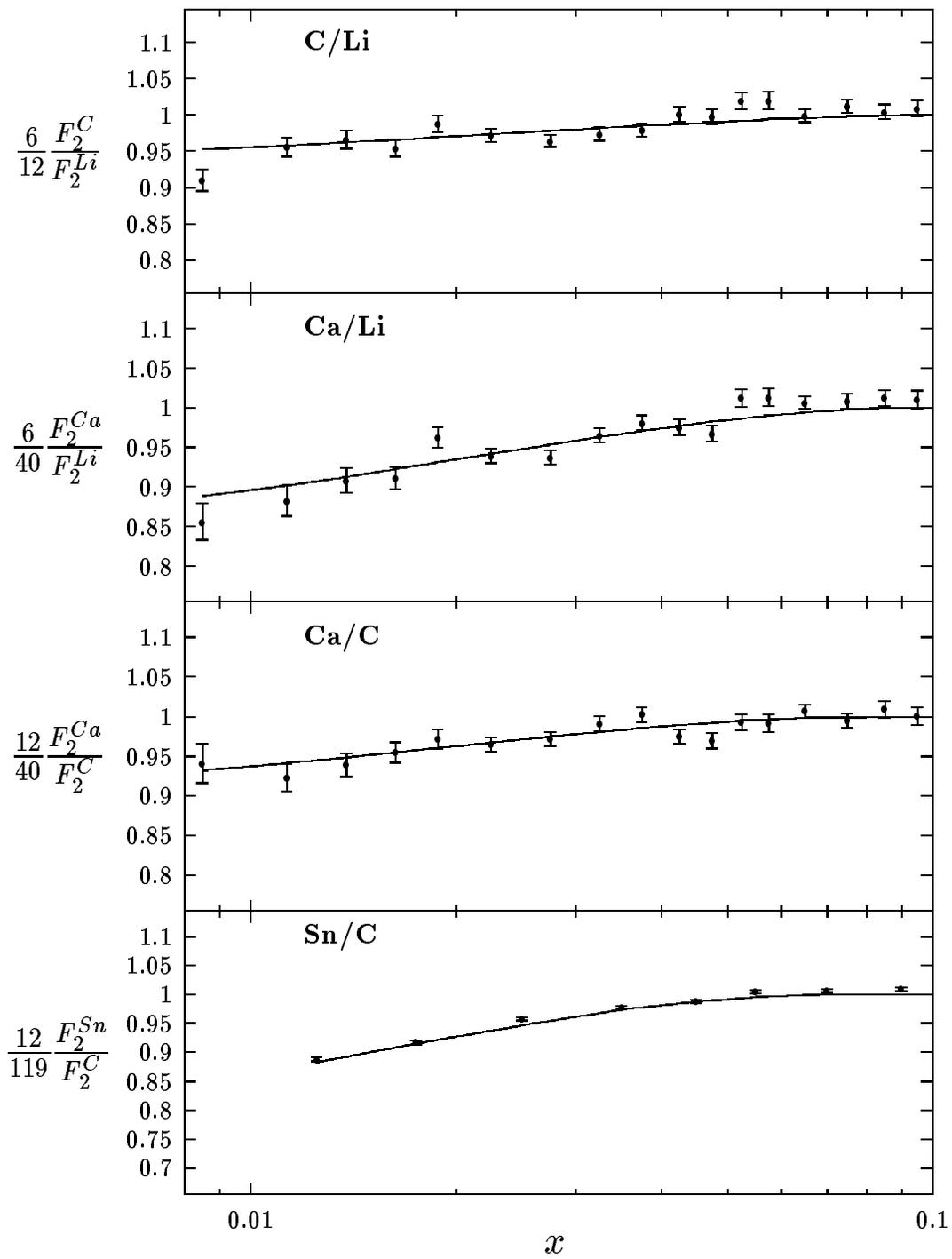
\item{\bf Fig. 4 :} The ratios $(A_1/A_2) F_2^{A_2}/F_2^{A_1}$ computed from
     eq. (3) for different values of $x$. The experimental points are from
     refs. [17] and [18]. The values of $Q^2$ are different for different $x$
     values [17, 18].

\vskip 3 truemm
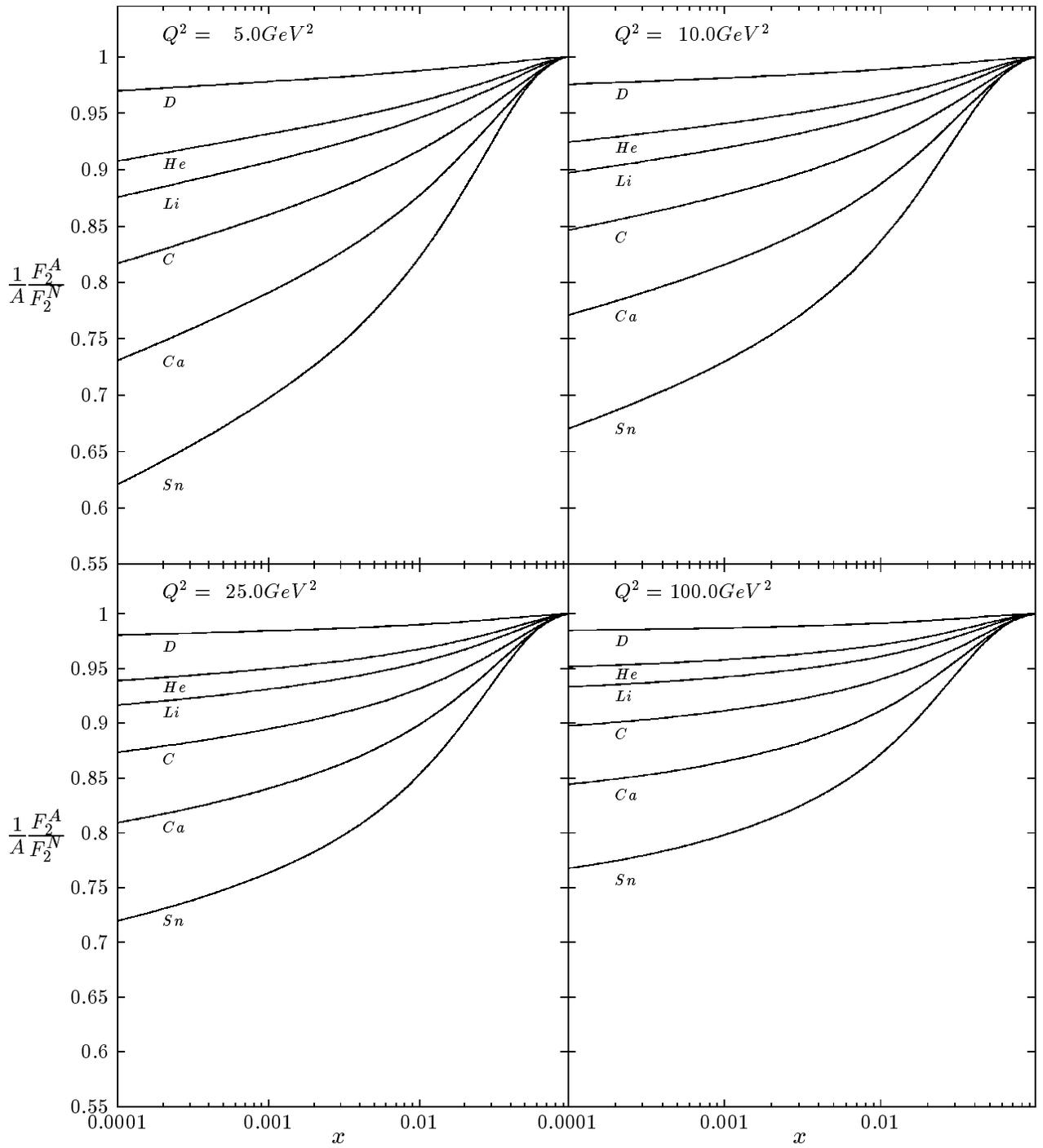
\item{\bf Fig. 5 :} The ratios $(1/A) F_2^{A}/F_2^{N}$ computed from eq. (3)
     for different values of $x$ in the small $x$ region, at fixed values of
     $Q^2$.

\vskip 3 truemm
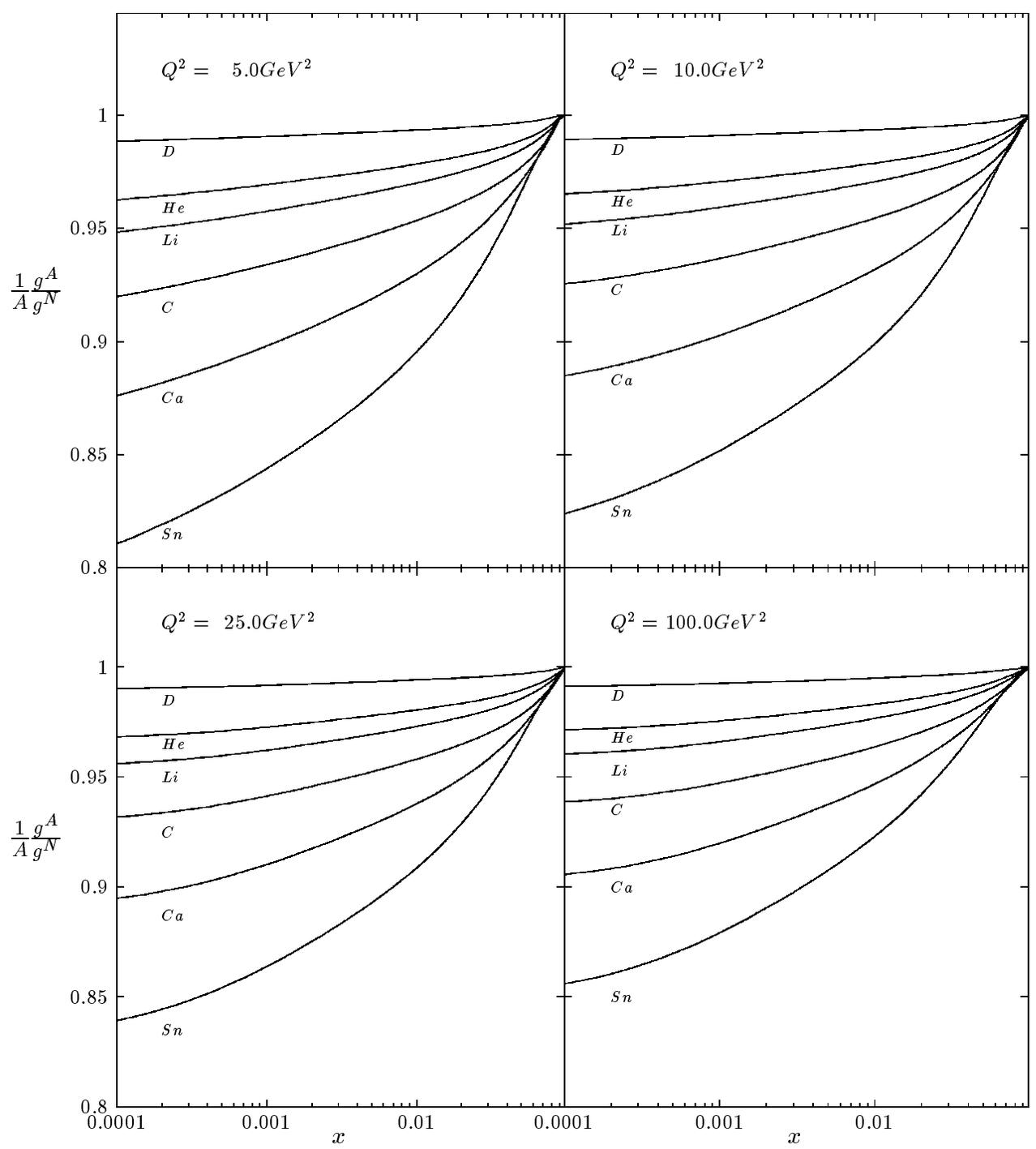
\item{\bf Fig. 6 :} The ratios $(1/A_2) g^A/g^N$ of gluon distribution
     functions computed for different values of $x$ in the low $x$ region, at
     fixed values of $Q^2$.

\vskip 3 truemm
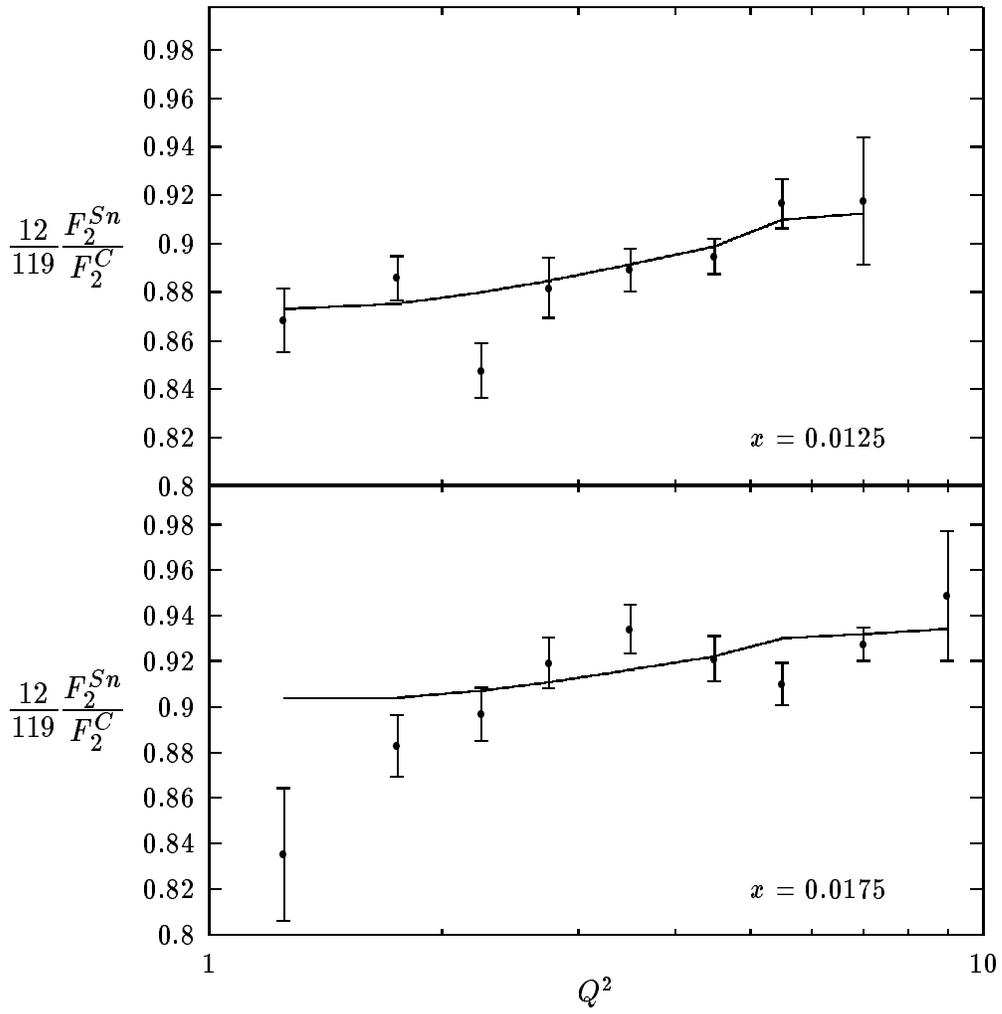
\item{\bf Fig. 7 :} The ratio $(12/119) F_2^{S_n}/F_2^C$ computed from eq. (3)
     for different values of $Q^2$, at two fixed values of $x$. The data points
     are from ref. [18]. \par
}

\bye